\documentclass[aps,prl,twocolumn,superscriptaddress,english]{revtex4}
\usepackage{amssymb}
\usepackage{graphicx}
\usepackage{amsmath}
\usepackage{soul}
\usepackage{amsthm}
\usepackage{amsfonts}
\usepackage[T1]{fontenc}
\usepackage[latin9]{inputenc}
\usepackage{array}
\usepackage{multirow}
\usepackage{color}
\usepackage{esint}
\usepackage{bm}
\usepackage{color}
\usepackage{bbm}
\usepackage{hyperref}
\usepackage{babel}
\usepackage{titlesec}

\begin{document}

\global\long\def\id{\mathbbm{1}}
\global\long\def\ui{\mathbbm{i}}
\global\long\def\ud{\mathrm{d}}

\title{Dissipation induced extended-localized transition}
\author{Yaru Liu}
\affiliation{Shenzhen Key Laboratory of Ultraintense Laser and Advanced Material Technology,
	Center for Intense Laser Application Technology, and College of Engineering Physics,
	Shenzhen Technology University, Shenzhen 518118, China}
\affiliation{Department of Physics, Renmin University of China, Beijing 100872, China}
\author{Zeqing Wang}
\affiliation{Department of Physics, Renmin University of China, Beijing 100872, China}
\author{Chao Yang}
\affiliation{Shenzhen Institute for Quantum Science and Engineering,
Southern University of Science and Technology, Shenzhen 518055, China}
\affiliation{International Quantum Academy, Shenzhen 518048, China}
\affiliation{Guangdong Provincial Key Laboratory of Quantum Science and Engineering, Southern University of Science and Technology, Shenzhen 518055, China}
\author{Jianwen Jie}
\thanks{Corresponding author: Jianwen.Jie1990@gmail.com}
\affiliation{Shenzhen Key Laboratory of Ultraintense Laser and Advanced Material Technology,
	Center for Intense Laser Application Technology, and College of Engineering Physics,
	Shenzhen Technology University, Shenzhen 518118, China}
\author{Yucheng Wang}
\thanks{Corresponding author: wangyc3@sustech.edu.cn}
\affiliation{Shenzhen Institute for Quantum Science and Engineering,
	Southern University of Science and Technology, Shenzhen 518055, China}
\affiliation{International Quantum Academy, Shenzhen 518048, China}
\affiliation{Guangdong Provincial Key Laboratory of Quantum Science and Engineering, Southern University of Science and Technology, Shenzhen 518055, China}
\begin{abstract}
Mobility edge (ME), representing the critical energy that distinguishes between extended and localized states, is a key concept in understanding the transition between extended (metallic) and localized (insulating) states in disordered and quasiperiodic systems. Here we explore the impact of dissipation on a quasiperiodic system featuring MEs 
by calculating steady-state density matrix and analyzing quench dynamics with sudden introduction of dissipation, 
and demonstrate that dissipation can lead the system into specific states predominantly characterized by either extended or localized states, irrespective of the initial state. Our results establish the use of dissipation as a new avenue for inducing transitions between extended and localized states, and for manipulating dynamic behaviors of particles. 
\end{abstract}
\maketitle

{\em Introduction.---} The investigation of electronic transport properties lies at the heart in condensed-matter physics~\cite{Wiersma2009}. Disorder is ubiquitous, and its crucial impact on transport properties was unveiled through Anderson localization (AL)~\cite{Anderson1958,RMP1}. In three-dimensional systems with substantial disorder, a transition can occur from the extended phase to the localized phase. Near this transition point, mobility edges (MEs) may emerge, defining the critical energy that distinguishes extended states from localized ones~\cite{RMP1,RMP2,Kramer1993}. ME is a vital focus in studying disordered materials and helps in understanding a material's conductivity and electronic properties. The extended (metal)-localized (insulator) transition can be induced by altering the position of the Fermi energy across the ME.
In addition to random disorder, quasiperiodic potentials can also induce the extended-localized transition (ELT)~\cite{Roati2008,Soukoulis1981,DasSarma1988,Biddle2009,XLi2017,HYao2019,Ganeshan2015,Wang1,XCZhou,Wang2022,TLiu,Longhi,SChen,Ribeiro,Bloch4,An2018,JiasT}, resulting in distinct physical phenomena different from that of disordered potentials. For instance, in one-dimensional (1D) quasiperiodic systems, the ELT and MEs can exist, whereas in disordered systems, these phenomena are expected to occur in dimensions higher than two according to scaling theory~\cite{Abrahams1979}. 
 
With advancements in non-Hermitian physics and the manipulation of both dissipation and quantum coherence in experimental settings, recent years have witnessed a growing interest in studying dissipative open quantum systems. Dissipation can profoundly change the properties of quantum systems, leading to various types of phase transitions~\cite{TProsen2008,Mebrahtu,Medvedyeva,Shastri2020,Soriente,Yamamoto,WNie,Kawabata,Brunelli,LingNaWu,Kuzmin}. 
The impact of dissipation on the localization and transport properties in disordered~\cite{Gurvitz2000,Nowak2012,Rayanov2013,Yamilov2012,Mujumdar2020,Weidemann2021,Huse2015,Yusipov,Yusipov2,Longhi2023} and quasiperiodic systems~\cite{Purkayastha2017,Saha2022,Lacerda2021,Goold0,Dwiputra2021,Goold1,Goold2} has also garnered widespread attention. It has been observed that dephasing noise can transform other forms of transport behavior into diffusive~\cite{Saha2022,Lacerda2021,Goold0}, thus, for subdiffusive and localized systems, dephasing can destroy AL and enhance transport. When considering the presence of MEs, the coupling of the system boundaries to baths may significantly increase environment-assisted quantum transport~\cite{Dwiputra2021} and energy current rectification~\cite{Goold1}.

Recently, Yusipov et al. applied a dissipative operator (given by Eq. (\ref{jumpoperator}) below) to a disordered system and discovered that it can drive AL into a stable state that retains its localized properties without being destroyed~\cite{Yusipov,Yusipov2}. In this letter, we investigate the impact of such dissipation on a 1D quasiperiodic system with MEs and find that the dissipation can drive the system into specific states, which may be extended or localized, regardless of the initial state. This reveals that the dissipation can induce the transition between extended and localized states. Such effects are not achievable with other types of dissipation, such as dephasing, energy decay, particle number decay, and so on. Remarkably, the ELT here does not necessitate the change in either disorder strength or particle density, both of which are believed to be the ways for altering the properties of localization through shifting the relative positions of the Fermi energy and the ME. Thus, the combination of such dissipation and MEs provides a new approach to induce ELT and manipulate a system's transport properties.

{\em Model.--- } We consider a dissipative system whose density matrix $\rho$ follows the Lindblad master equation~\cite{GLindblad,HPBreuer}, 
\begin{align}
	\label{lindblad}
	\frac{d\rho(t)}{dt} =\mathcal{L} [\rho(t)]= -i\left[H, \rho(t)\right] +\mathcal{D}[ \rho(t)].
\end{align}
where $\mathcal{L}$ is referred to as the Lindbladian, with its dissipative component denoted as,
\begin{eqnarray}\label{Dj}
	\mathcal{D}[\rho(t)] =  \Gamma\sum_j\left(O_{j}\rho O_{j}^{\dagger} -1/2\{ O_{j}^{\dagger} O_{j} , \rho \}\right),
\end{eqnarray}
which contains a set of jump operators $O_{j}$, all with the same strength $\Gamma$ here. 
Assuming that $\mathcal{L}$ is time-independent, we can express $\rho(t)=e^{\mathcal{L}t}\rho(0)$. One can define the steady state as $\rho_{ss}=\lim_{t\rightarrow\infty}\rho(t)$, which corresponds to the eigenstate of the Lindblad generator with zero eigenvalue, i.e., $\mathcal{L}[\rho_{ss}] = 0$.

The Hamiltonian we consider in Eq.~(\ref{lindblad}) is denoted as
\begin{eqnarray}\label{Hamiltonian}
	H=J\sum_{j=1}\left(c_{j}^{\dagger}c_{j+1}+\text{h.c.}\right)+2\sum_{j=1}V_{j}n_{j},
\end{eqnarray}
where $c_{j}$ and $n_{j}=c_{j}^{\dag}c_{j}$ are respectively the annihilation operator and local number operator at site $j$, and $J$ is the nearest neighbor hopping coefficient.
The local potential $V_{j}$ takes $V\cos\left[2\pi\omega j+\theta\right]$ for even sites and vanishes for odd sites, where $\omega$ is an irrational number, $V$ and $\theta$ are potential amplitude and phase offset, respectively. 
The model is referred to as the 1D quasiperiodic mosaic model, which features two MEs at $E_{c}=\pm J/V$~\cite{Wang1}. These two MEs divide the energy spectrum into three regions, including one extended region ($-J/V< E <J/V$) and two localized regions corresponding to $E > J/V$ and $E <-J/V$, respectively. This model has been recently realized, and the MEs have been detected~\cite{Gao2023}. Without loss of generality, we set $J=1$, $V=1$, $\theta=0$, and $\omega=(\sqrt{5}-1)/2$. Unless otherwise stated, we use open boundary conditions in subsequent calculations.

The jump operator considered in Eq. (\ref{Dj}) is given by~\cite{Yusipov,Yusipov2,explainsign,PZoller1,PZoller2,PZoller3,PZoller4,Marcos} 
\begin{eqnarray}\label{jumpoperator}
	O_{j}=(c^{\dagger}_{j}+e^{i\alpha}c_{j+l}^{\dagger})(c_{j}-e^{i\alpha}c_{j+l}),
\end{eqnarray}
which acts on a pair of sites $j$ and $j+l$. This jump operator does not alter the system's particle number, but it does change the relative phase between the pair of sites with the distance $l$. For example, this operator synchronizes them from an out-of-phase (in-phase) mode to an in-phase (out-of-phase) mode when the dissipative phase $\alpha$ is set to $0$ ($\pi$). Such property is important for attaining the intended extended or localized stationary states, as will become evident below.

\begin{figure}
	\centering
	\includegraphics[width=0.49\textwidth]{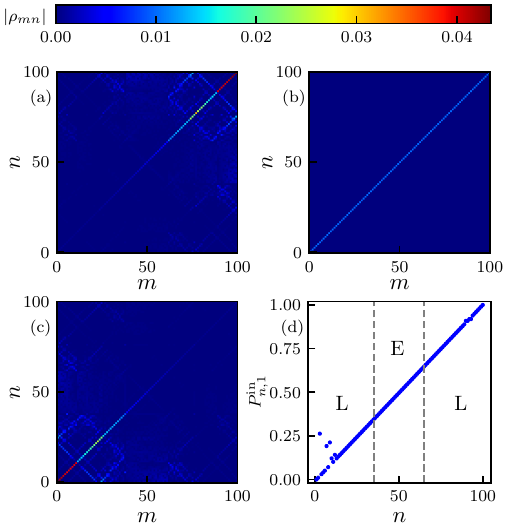}
	\caption{\label{01} The absolute values of the density matrix elements for steady states with the dissipative phases (a) $\alpha=0$, (b) $\alpha=\pi/2$, and (c) $\alpha=\pi$ in the eigenbasis of the Hamiltonian $H$. (d) The proportion of in-phase site pairs for each eigenstate. Dashed lines mark the MEs, separating eigenstates into localized (L), extended (E), and localized (L) regions as eigenvalues increase. Here we take $l=1$, $L=100$ and $\Gamma=1$. }
\end{figure}

{\em Dissipation induced ELT.---} We begin by examining the jump operators in Eq. (\ref{jumpoperator}) for $l=1$ and analyzing the properties of the stationary solution $\rho_{ss}$ in the eigenbasis of the Hamiltonian $H$, that is, $\rho_{mn}=\langle \psi_m|\rho_{ss}|\psi_n\rangle$, where $|\psi_m\rangle$ and $|\psi_n\rangle$ are the eigenstates of the Hamiltonian $H$. Fig.~\ref{01}(a-c) illustrate the transition of the system's steady state from being predominantly composed of high-energy localized states to predominantly composed of low-energy localized states as the dissipation phase is varied from $\alpha=0$ to $\alpha=\pi$. The steady state here is independent of the initial state. 
This implies that if the initial state is within the extended region located in the middle of the energy spectrum, when the jump operator with $\alpha=0$ or $\alpha=\pi$ is introduced, the state will ultimately predominantly concentrate on localized regions.
	
The composition of the steady states can be understood by analyzing the relative phases of neighboring lattice sites. For an arbitrary $n$-th eigenstate $|\psi_n\rangle=\sum_{j}^L\psi_{n,j}c_{j}^{\dagger}|\varnothing\rangle$ with $L$ being the system size, we can calculate the phase difference $\Delta\phi^n_{i,l}$ between the $i$-th and the $(i+l)$-th lattice points as $\Delta\phi^n_{j,l}=\arg(\psi_{n,j})-\arg(\psi_{n,j+l})$. If $\Delta\phi^n_{j,l}=0$, it implies that they are in phase. Therefore, we can calculate the number of in-phase site pairs $N_{n,l}^{\text{in}}$ with distance $l$, and its proportion $P_{n,l}^{\text{in}}=N_{n,l}^{\text{in}}/N_t$, where $N_t=L-l$ represents the total number of site pairs. Fig. \ref{01}(d) illustrates that eigenstate with higher (lower) energy tends to have larger (smaller) $P_{n,1}^{\text{in}}$~\cite{SM}, explaining why the steady state predominantly concentrates on high-energy (low-energy) localized eigenstates for $\alpha=0~(\alpha=\pi)$. 

\begin{figure}
	\centering
	\includegraphics[width=0.49\textwidth]{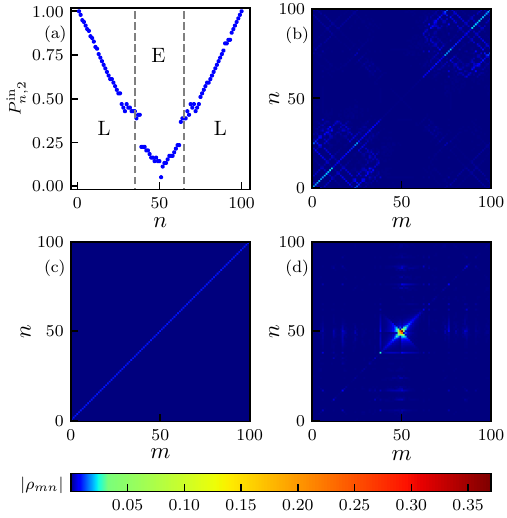}
	\caption{\label{02} (a) The proportion of in-phase lattice site pairs for each eigenstate, with the mobility edges marked by the dashed lines. Like in Fig.~\ref{01}, the dashed lines here represent MEs.  The absolute values of the steady-state density matrix elements with the dissipative phases (b) $\alpha=0$, (c) $\alpha=\pi/2$, and (d) $\alpha=\pi$ in the eigenbasis of Hamiltonian $H$. Other parameters are $l=2$, $L=100$ and $\Gamma=1$.}
\end{figure}

We further investigate the effect of the jump operators in Eq. (\ref{jumpoperator}) with $l=2$. We first calculate the proportion of in-phase lattice site pairs $P_{n,2}^{\text{in}}$, and find that it exhibits a $V$-shaped pattern~[Fig. \ref{02}(a)]. The localized states on both sides of the energy spectrum tend to exhibit more in-phase site pairs, while the extended states in the middle of the spectrum tend to have more out-of-phase site pairs~\cite{SM}. Therefore, by appropriately choosing the dissipation phase $\alpha$, it is possible to control whether the system's steady state is predominantly composed of localized or extended eigenstates, as shown in Fig. \ref{02}(b-d). When the dissipative phase is $\alpha=0$ [Fig. \ref{02}(b)], the system is expected to reach a steady state predominantly composed of the states associated with in-phase site pairs, thus primarily concentrating on the localized eigenstates in both higher-energy and lower-energy regions. Conversely, when $\alpha=\pi$ [Fig. \ref{02}(d)], the system is anticipated to attain a steady state mainly composed of the states linked to out-of-phase site pairs, favoring the dominance of extended eigenstates in the mid-energy regions. When $\alpha=\pi/2$ [Fig. \ref{02}(c) and Fig. \ref{01}(b)], the dissipation operator becomes Hermitian, leading to the system reaching the maximally-mixed state $(\rho_{ss})_{mn}=\delta_{mn}/L$ as its steady state. According to the diagonal elements of the density matrix $\rho_{mn}$, for arbitrary $\alpha$, we can determine the fractions of localized and extended eigenstates in steady states, i.e., $P_{l}=\sum_{i}\rho_{ii}$ ($P_e=\sum_j\rho_{jj}$), where $i (j)$ represents the index of the extended (localized) eigenstates $|\psi_i\rangle$ ($|\psi_j\rangle$). When $\alpha$ is tuned from $0$ to $\pi$, the system's steady state shows the transition from being dominated by localized eigenstates to being dominated by extended eigenstates [Fig. \ref{03}(a)], which indicates that dissipation can be used to manipulate the ELT. In Supplementary Material~\cite{SM}, we investigated the variations of $P_l$ and $P_e$ with system size. It is observed that when $\alpha=0$, $P_e$ and $P_l$ show minimal changes with size, while for $\alpha=\pi$, as the size increases, $P_l$ ($P_e$) gradually tends towards $0$ ($1$).

\begin{figure}
	\hspace*{-0.1cm}
	\centering
	\includegraphics[width=0.485\textwidth]{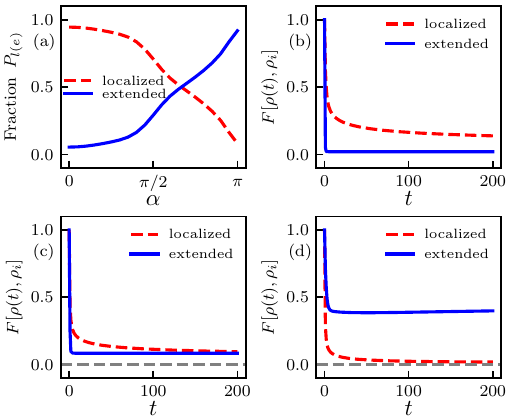}
	\caption{\label{03} (a) The fractions of localized eigenstates ($P_l$) and extended eigenstates ($P_e$) in steady states as a function of the dissipative phase $\alpha$. Evolution of the fidelity defined in Eq. (\ref{fed}) after a sudden introduction of dissipation with the strength $\Gamma=1$ and the phases (b) $\alpha=0~$, (c) $\alpha=\pi/2$, and (d) $\alpha=\pi$. The initial states are set as follows: the system's ground state, which is localized (red dashed line), and the state corresponding to the eigenvalue situated in the center of the energy spectrum of the Hamiltonian $H$, characterized as extended (blue solid line). Here, $L=144$ and $l=2$. }
\end{figure}

We further examine this transition from a dynamical perspective. We prepare a localized or extended eigenstate as the initial state and introduce dissipation with $l=2$ at $t=0$. We then calculate the fidelity, which represents the overlap between the time-evolved state $\rho(t)$ and the initial state $\rho_{i}$, denoted as~\cite{Nielsen2000,Zanardi2007}
	\begin{eqnarray}\label{fed}
	F[\rho(t),\rho_{i}]= Tr[\sqrt{\rho(t)^{1/2}\rho_{i}\rho(t)^{1/2}}].
	\end{eqnarray}
At $\alpha=0$ [Fig. \ref{03}(b)], the fidelity rapidly approaches zero when the initial state is extended, indicating that the structure of the initial state has been completely modified. Conversely, the fidelity tends to a non-zero value when the initial state is localized, suggesting that certain characteristics of the initial state are preserved. This is because the steady state is primarily composed of localized states, i.e., $P_l\gg P_e$. When $\alpha=\pi/2$ [Fig. \ref{03}(c)], the steady state contains extended and localized states, resulting in non-zero fidelity for both. Finally, for $\alpha=\pi$ [Fig. \ref{03}(d)], the steady state primarily consists of extended states. Therefore, when the initial state is an extended state, the fidelity is not equal to zero, but it tends to zero when the initial state is localized.
		
\begin{figure}
		\hspace*{-0.1cm}
		\centering
		\includegraphics[width=0.485\textwidth]{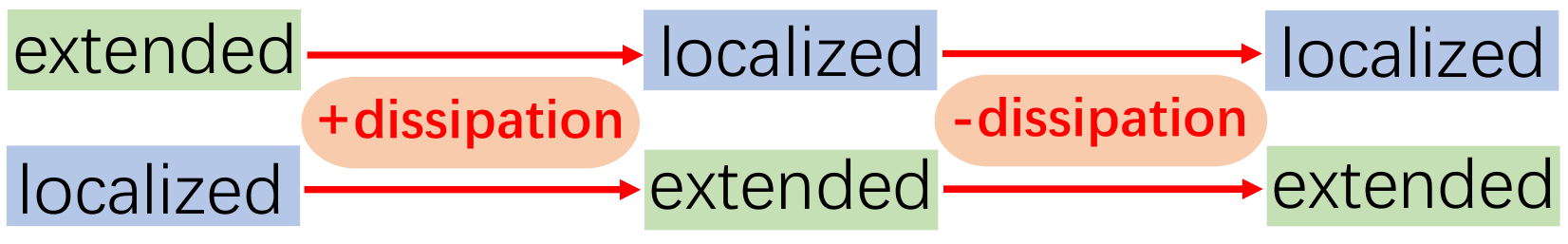}
		\caption{\label{04} Schematic illustrating transitions between extended and localized states manipulated by dissipation. }
	\end{figure}

We have revealed that an extended (localized) state can be guided towards a steady state primarily composed of localized (extended) states by applying dissipation. Furthermore, even after removing the dissipation once the system has reached a steady state, its properties continue to persist~\cite{explain0}. This differs from the majority of previous studies on the influence of dissipation on AL, where dissipation disrupts localization, but removing it leads the disordered system back to localization~\cite{explain1}. By introducing a period of dissipation and subsequently removing it, the parameters in the Hamiltonian remain unchanged, but the dynamical properties undergo a profound transformation, as shown in Fig. \ref{04}. Consequently, dissipation provides a means to manipulate transitions between localized and extended states. 
	
\begin{figure}
	\centering
	\includegraphics[width=0.5\textwidth]{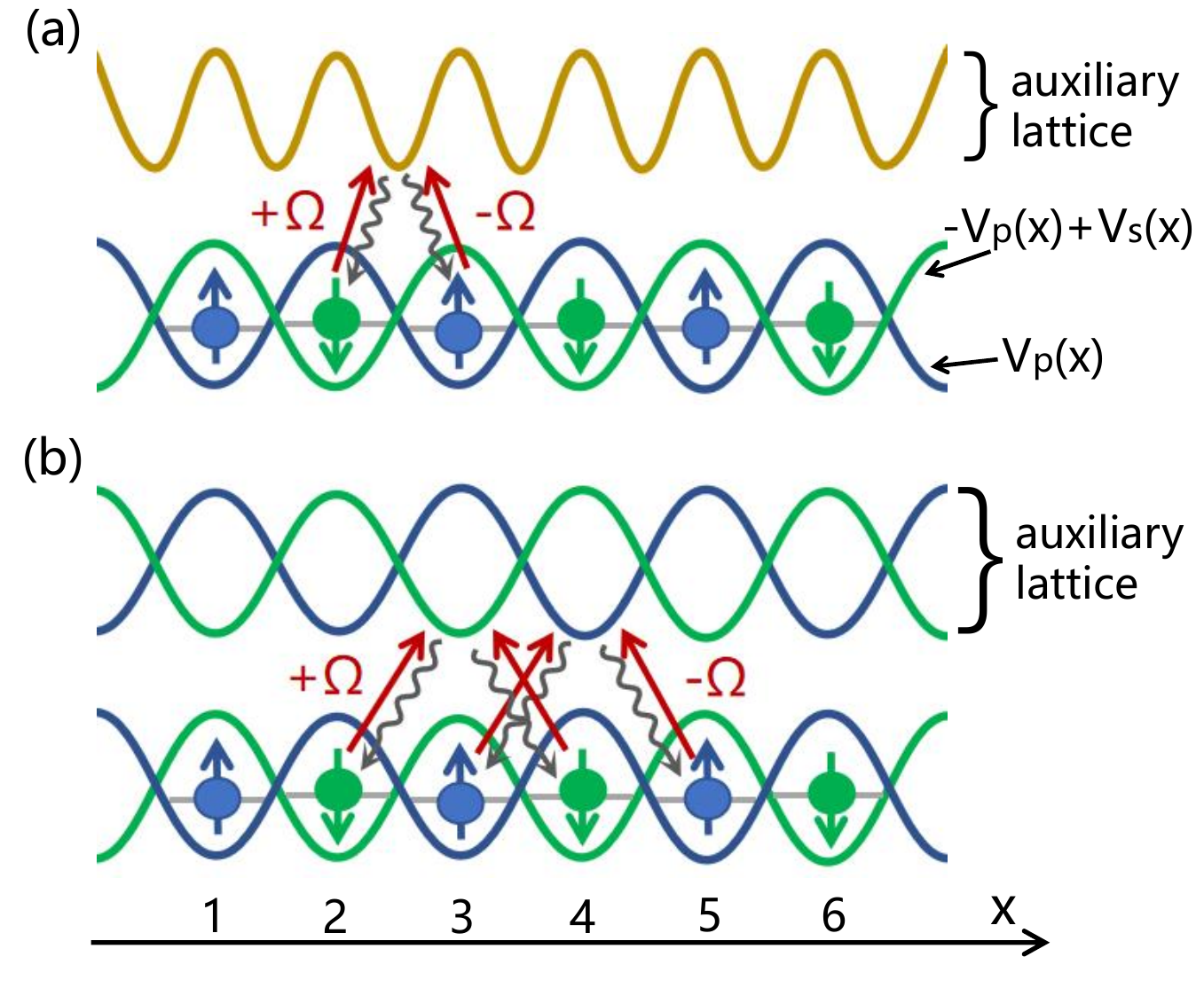}
	\caption{\label{05}
		 Schematic realization of the dissipative process of the form $(c^{\dagger}_j+c^{\dagger}_{j+l})(c_j-c_{j+l})$ with (a) $l=1$ and (b) $l=2$. The lower (upper) lattice correspond to the physical (auxiliary) lattice. Atoms in the lower sites are coupled to the auxiliary sites in between with opposite Rabi frequency $\pm \Omega$, and then spontaneously decay back to the lower sites. }
\end{figure}

{\em Experimental realization.---} Ref.~\cite{Wang1} has proposed realizing the quasiperiodic mosaic model based on optical Raman lattices~\cite{XJLiu,BSong,LZhang,ZWu}. They constructed a spin-1/2 system using two hyperfine states $|\!\uparrow\rangle=|F_1,m_{F1}\rangle$ and $|\!\downarrow\rangle=|F_2,m_{F2}\rangle$, and mapped the spin-up (spin-down) lattice sites to the odd (even) sites [see Fig.~\ref{05}]. They applied the Raman coupling via a standing wave field and a plane wave to generate the spin-dependent primary lattice $V_p(x)\sigma_z=V_1\cos(2k_px+\phi_p)\sigma_z$, and three standing wave fields together to generate a secondary quasiperiodic lattice $V_s(x)=V_2\cos(2k_sx+\phi_s)$ only for spin-down atoms. As the depth of the primary lattice significantly exceeds that of the secondary lattice ($V_1\gg V_2$), we will proceed to discuss the realization of the dissipation operator below, with the secondary lattice's impact considered negligible. We will primarily focus on discussing the realization of the case with $\alpha=0$ by introducing the auxiliary lattice~\cite{PZoller1,PZoller2,PZoller3,PZoller4}. An arbitrary phase $\alpha$ can be achieved, for instance, through an array of resonators coupled by superconducting qubits~\cite{Yusipov,Marcos}. 

Fig.~\ref{05}(a) shows the realization of the pairwise dissipator with $l=1$. By coherently coupling two nearly degenerate levels in the system to an auxiliary site in between with antisymmetric Rabi frequencies $\pm \Omega$, which can be obtained by controlling the wavelength of the driving laser to match that of the primary lattice, one can achieve the annihilation part of the dissipative operator. Decay back to the lower sites occurs through spontaneous emission, and this process is isotropic, leading to the form of the creation operator being symmetric~\cite{PZoller1,PZoller2,PZoller3,PZoller4}. In this way, one can realize the jump operator of the form $(c^{\dagger}_j+c^{\dagger}_{j+1})(c_j-c_{j+1})$.

Similarly, the case for $l=2$ can also be realized, as shown in Fig.~\ref{05}(b). However, there are several key differences with the case of $l=1$. (1) The auxiliary lattice must be spin-dependent, necessitating the use of two hyperfine states $|\!\uparrow\rangle=|F'_1,m'_{F1}\rangle$ and $|\!\downarrow\rangle=|F'_2,m'_{F2}\rangle$ to construct a spin-1/2 lattice. It is essential to ensure that they satisfy the condition $m_{F1}-m'_{F1}=m_{F2}-m'_{F2}$. (2) It requires a phase difference of $\pi$ with the primary lattice, namely that the odd (even) sites correspond to spin-down (spin-up) sites in the auxiliary lattice. (3) By manipulating the polarization of the driving laser, one can attain the specific coupling that spin-up (spin-down) states in the primary lattice only couple with spin-up (spin-down) states in the auxiliary lattice~\cite{Arimondo1977,Weis2002}. For example, if $m_{F1}-m'_{F1}=0$, the driving laser needs to be $\pi$-polarized~\cite{Jenkins2022,Chen2022}. Moreover, to obtain a $\pi$ phase shift in the effective Rabi frequency $\Omega$ from one spin-up (spin-down) site to the next spin-up (spin-down), 
one need to set the driving laser's wavelength to be twice that of the standing wave laser generating the primary lattice.

By comparing the diffusion~\cite{Ketzmerick,Larcher,YuWang,Shimasaki} or transport behaviors~\cite{Wang2022,Saha,JPBrantut,CCChien} of atoms before and after the introduction of dissipation, one can obtain extended and localized information about the initial and final states, thereby detecting the ELT caused by dissipation.

{\em Conclusion and Discussion.---} We have investigated the influence of dissipation on the 1D quasiperiodic mosaic model, which possesses exact MEs, and proposed its experimental realization. By calculating the distribution of the steady-state density matrix and the characteristics of quench dynamics, we revealed that dissipation can drive the system into specific states primarily composed of either extended or localized states, regardless of the initial states. Hence, dissipation can be utilized to induce transitions between extended and localized states, thereby enabling the manipulation of particle transport behaviors. In addition to its applications in condensed matter physics, our results also have potential applications in quantum simulation, specifically in controlling the dynamical behavior of particles and preparing desired states. Give a specific example, when simulating systems with MEs using cold atoms, preparing atoms near the ground state is challenging when they are localized, but introducing dissipation can aid in achieving this goal.

The manipulation of the ELT is expected to be common in various systems featuring MEs~\cite{SM}. When considering the impact of this dissipation on the anomalous MEs separating critical states from localized or extended ones, it is found that dissipation can induce critical-localized or critical-extended transitions (see supplementary materials~\cite{SM}). Our results pose several interesting issues. How such dissipation affects a many-body system with MEs? Can the dissipation be employed to manipulate the transitions between thermalized states and many-body localized states? Can the similar ETL exist in a three dimensional dissipative disordered system with MEs? Further, the dissipative operator in Eq. (\ref{jumpoperator}) utilizes the phase distribution characteristics of different states in the energy spectrum to select specific states as steady states. This provides a new perspective, suggesting that we can also explore other distinguishing features to construct experimentally feasible dissipative operators for the purpose of selecting different states.

\begin{acknowledgments}
We thank Long Zhang, Xin-Chi Zhou, Shi Yu, and Bing Yang for valuable discussions. This work was supported by National Key R\&D Program of China under Grant No.2022YFA1405800, the National Natural Science Foundation of China (Grant No. 12104210, Grant No. 12104205) and the Natural Science Foundation of Top Talent of SZTU(GDRC202202). Y. Liu acknowledges support from the Fundamental Research Funds for the Central Universities, and the Research Funds of Renmin University of China (22XNH100).
\end{acknowledgments}
	

\global\long\def\id{\mathbbm{1}}
\global\long\def\ui{\mathbbm{i}}
\global\long\def\ud{\mathrm{d}}

\setcounter{equation}{0} \setcounter{figure}{0}
\setcounter{table}{0} 
\renewcommand{\theparagraph}{\bf}
\renewcommand{\thefigure}{S\arabic{figure}}
\renewcommand{\theequation}{S\arabic{equation}}

\onecolumngrid
\flushbottom
\newpage
\section*{\large Supplementary Material:\\Dissipation induced extended-localized transition}
In the Supplementary Materials, we first provide an example to understand the distribution of phase differences in the main text, and analyze the fractions of extended and localized states changing with size. Then we discuss the impact of dissipation on three systems with exact mobility edges (MEs). The first system has the traditional ME that separates extended states from localized ones, allowing dissipation to induce the transitions between extended and localized states. The second and third systems feature the anomalous MEs that separate localized and extended states from critical states, enabling dissipation to induce the critical-localized and critical-extended transitions. Finally, we consider the impact of dissipation on the Dyson model.

\subsection{I. A simple example to understand the distribution of phase differences}
\begin{figure}[h]
	\centering
	\includegraphics[width=0.25\textwidth]{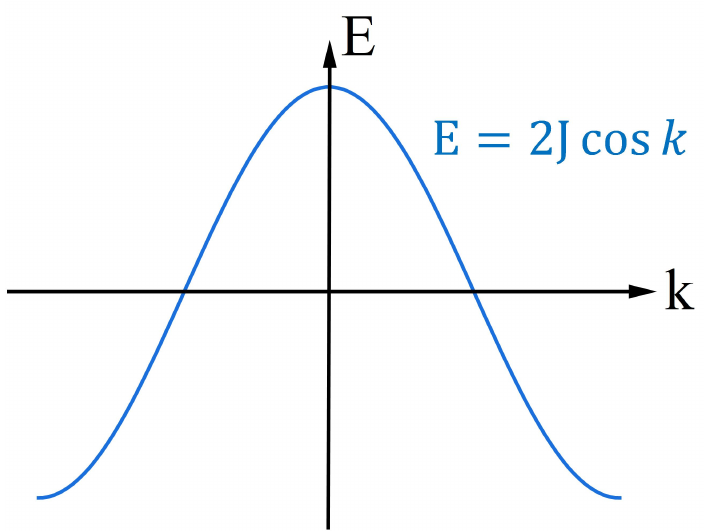}
	\caption{\label{S3}
		The eigenvalue distribution of $H_0$: $E=2J\cos(k)$.}
\end{figure}

In this section, we provide an example to illustrate the generality of the results regarding the proportion of in-phase states in the main text. We consider a simple Hamiltonian with only a hopping term, i.e., $H_0=J\sum_j(c_j^{\dagger}c_{j+1}+h.c.)$. 
Its wave function is given by $e^{ikj}$, with $k=2\pi n/L (n\in(-L/2,L/2])$. It is easy to obtain the phase difference between neighboring lattice points $\Delta\phi=k(j+1)-kj=k$, as well as the phase difference for next-nearest neighbors $\Delta\phi=2k$.
Its eigenvalues are given by $E=2J\cos(k)$, as shown in Fig. \ref{S3}. We can observe that at the top of the band where $k=0$, it implies that the phase differences for both neighboring and next-nearest neighboring lattice points are $0$, indicating they are in-phase. At the bottom of the band where $k=\pi$, it implies that the neighboring lattice points are out-of-phase, and the next-nearest neighboring lattice points are in-phase. For the middle of the spectrum at $k=\pi/2$, the next-nearest neighboring lattice points are out-of-phase. This is consistent with our manuscript, where the minimum and maximum values of $P_{n,1}^{\text{in}}$ are located at the low and high ends of the energy spectrum, respectively (Fig.1(d) in the main text). The maximum value of $P_{n,2}^{\text{in}}$ is situated at both ends of the energy spectrum, while the minimum value is located in the middle of the energy spectrum (Fig.2(a) in the main text). Therefore, our phase analysis results can be applied to many models, indicating that we can use the dissipative operators mentioned to achieve state selection in various systems, triggering transitions between extended and localized states, and even causing some other transitions. 

\subsection{II. $P_l$ and $P_e$ change with the system size}
\begin{figure}[h]
	\centering
	\includegraphics[width=0.95\textwidth]{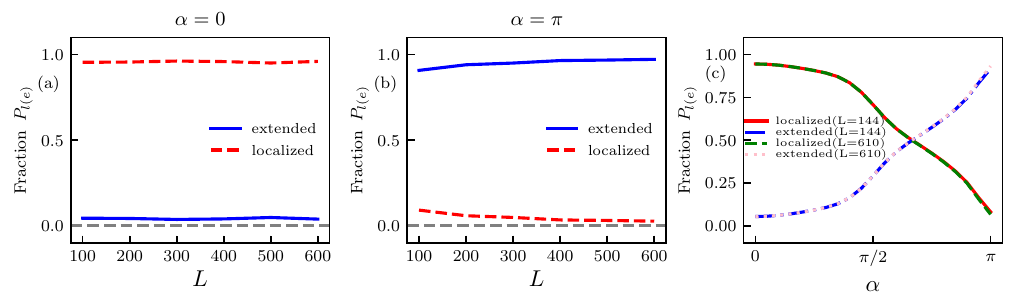}
	\caption{\label{S4}
		The fractions of localized eigenstates ($P_l$) and extended eigenstates ($P_e$) in steady states as a function of the system sizes for (a) $\alpha=0$ and (b) $\alpha=\pi$. (c) $P_l$ and $P_e$ as a function of the dissipative phase $\alpha$ for the system sizes $L=144$ and $L=610$. The other parameters are the same as those in Fig. 3(a) of the main text.}
\end{figure}

In Fig.3(a) in the main text, we plotted the fractions of extended and localized states as a function of $\alpha$. It can be observed that at $\alpha=0$, the fraction of localized states ($P_l$) is much greater than that of extended states ($P_e$), while at $\alpha=\pi$, $P_e$ is much greater than $P_l$. The Figs. \ref{S4} (a) and (b) respectively show the variations of $P_e$ and $P_l$ with size at $\alpha=0$ and $\alpha=\pi$. It can be seen that when $\alpha=0$, $P_e$ and $P_l$ are around $0.05$ and $0.95$ respectively, and they hardly change with the variation of size. When $\alpha=\pi$, with the increase in size, $P_l$ ($P_e$) gradually tends towards $0$ ($1$). Certainly, the magnitudes of $P_l$ and $P_e$ also depend on the model.

When the states to be selected are neither in the central region nor on the edges of the energy spectrum, perhaps considering intermediate values of $\alpha$ could achieve this purpose. Fig. \ref{S4}(c) illustrates the fractions of localized eigenstates ($P_l$) and extended eigenstates ($P_e$) in steady states as a function of the dissipative phase $\alpha$ for two different system sizes. We see that the lines with different sizes largely overlap, indicating that for any intermediate $\alpha$, $P_l$ and $P_e$ quickly converge to stable values. Additionally, zooming in on the region near $\alpha=\pi$, we can observe that $P_l$ and $P_e$ gradually approach $0$ and $1$.

\begin{figure}[h]
	\includegraphics[width=0.85\textwidth]{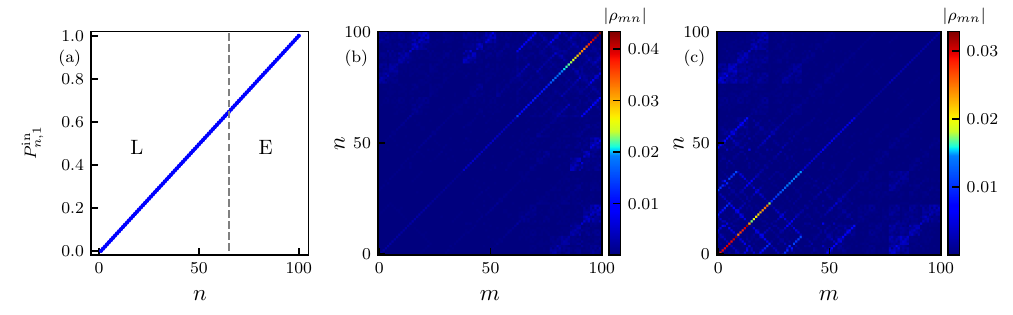}
	\caption{\label{S1} (a) The proportion of in-phase lattice site pairs for each eigenstate. Dashed lines are the MEs satisfying Eq. (\ref{MEs3}), separating eigenstates into localized (L) and extended (E) regions as eigenvalues increase.
		The absolute values of the density matrix elements for steady states with the dissipative phases (b) $\alpha=0$ and (c) $\alpha=\pi$ in the eigenbasis of Hamiltonian given in Eq. (\ref{ME2}).  Here we take $t_1=0.8$, $\lambda=0.825$, $\beta=-0.1$, $\delta=0$, $\omega=(\sqrt{5}-1)/2$, $L=100$ and $\Gamma=1$.}
\end{figure}

\subsection{III. The impact of dissipation on a system with only one ME}
We consider the following model~\cite{Ganeshan2015S}
\begin{equation}
	H_1= t_1\sum_{j=1}\left(c_{j}^{\dagger}c_{j+1}+\text{h.c.}\right)+2\lambda\sum_{j=1}\frac{\cos(2\pi\omega j+\delta)}{1-\beta\cos(2\pi\omega j+\delta)}n_{j},
	\label{ME2}
\end{equation}
where $\omega$ is an irrational number, $t_1$, $\lambda$, and $\beta$ ($\beta\in (-1,1)$) represent the hopping strength, the on-site modulation strength, and the deformation parameter, respectively. The exact expression of its ME is~\cite{Ganeshan2015S}
\begin{equation}
	E_{1c}=2sgn(\lambda)(|t_1|-|\lambda|)/\beta.
	\label{MEs3}
\end{equation}

We introduce the jump operators described by Eq. (4) in the main text with $l=1$ and study the distribution of the steady-state density matrix $\rho_{ss}$ in the eigenbasis of the Hamiltonian $H_1$. As in the main text, we set $\rho_0=\rho_{L+1}=0$ at the boundaries, where $L$ represents the system size. Fig. \ref{S1}(a) shows the $P_{n,1}^{\text{in}}=N_{n,1}^{\text{in}}/N_t$ for each eigenstate, where $N_{n,l}^{\text{in}}$ and $N_t$ are the number of in-phase site pairs and the total number of site pairs, respectively, with $N_t=L-1$. Similar to Fig. 1(d) in the main text, eigenstate with higher (lower) energy tends to have larger (smaller) $P_{n,1}^{\text{in}}$. However, the localized and extended properties of this system differ from those of the model in the main text. Here, there is only one ME that separates extended states from localized states. Thus, 
when the dissipative phase $\alpha=0$, the steady state predominantly concentrates on high-energy extended eigenstates [Fig. \ref{S1}(b)], while at $\alpha=\pi$, the steady state predominantly concentrates on low-energy localized eigenstates [Fig. \ref{S1}(c)]. For other models with MEs, we have also found that the jump operator described by Eq. (4) in the main text can manipulate the transition between extended and localized states. This demonstrates the generality of this manipulation method.

\subsection{IV. The impact of dissipation on the systems with anomalous MEs}
In addition to the traditional ME separating extended states from localized states, recent attention has been drawn to novel anomalous MEs that separate localized states from critical states~\cite{XCZhouS,TLiuS,Wang2022S}, as well as separating extended states from critical states~\cite{Wang2022S,DengXLS}. Our findings also apply to these new types of MEs.
\subsubsection {A. Dissipation induced critical-localized transition}
The model we examined, which possesses precise MEs that distinguish critical and localized states, is denoted as~\cite{XCZhouS}
\begin{equation}
	H_2=\sum_{j}(t_{j}a_{j}^{\dagger}a_{j+1}+\mathrm{h.c.})+\sum_{j}\lambda_{j}n_{j},\label{Ham2}
\end{equation}
where both the on-site potential $\lambda_{j}$ and the hopping coefficient $t_{j}$ are mosaic, with
\begin{equation}
	\{t_{j}, \lambda_{j}\}=\begin{cases}
		\{t_2,\ 2\lambda_0\cos[2\pi\omega(j-1)+\theta]\}, &  \mathrm{mod\,}(j,2)=1,\\
		2\lambda_0\cos(2\pi\omega j+\theta)\{1,\ 1\}, &  \mathrm{mod\,}(j,2)=0.
	\end{cases}\label{eq:mosaic}
\end{equation}
with $t_2$ and $\theta$ being hopping coefficient and phase offset, respectively. For convenience, we set $\lambda_0=1$, $\theta=0$, $\omega=(\sqrt{5}-1)/2$, and the MEs are given by
\begin{equation}\label{MEEE}
	E_c=\pm t_2.
\end{equation}

\begin{figure}[h]
	\includegraphics[width=0.85\textwidth]{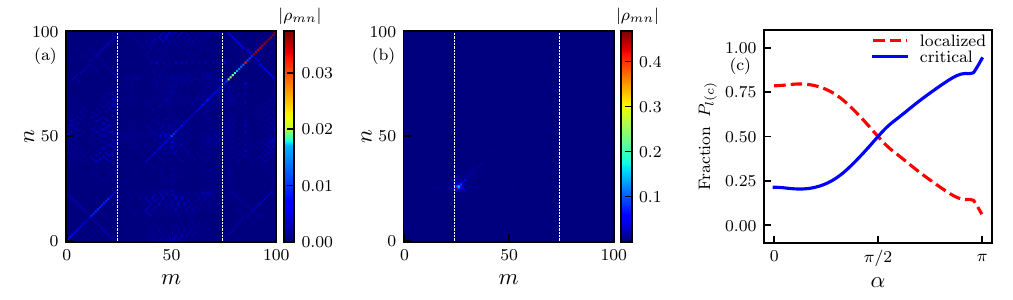}
	\caption{\label{S2}	The absolute values of the density matrix elements for steady states with the dissipative phases (a) $\alpha=0$ and (b) $\alpha=\pi$ in the eigenbasis of Hamiltonian given in Eq. (\ref{Ham2}).   Dashed lines are the MEs satisfying Eq. (\ref{MEEE}), separating localized states from critical ones. (c) The fraction of localized eigenstates ($P_l$) and critical eigenstates ($P_c$) in steady states versus $\alpha$, and the results are size-independent.
		Other parameters are $\lambda_0=1$, $t_2=1$, $\omega=(\sqrt{5}-1)/2$, $L=100$, $l=1$ and $\Gamma=1$.}
\end{figure}

Similarly, we introduce the identical dissipative operators with $l=1$ and set $\rho_0=\rho_{L+1}=0$ at the boundaries. Subsequently, we analyze the distribution of the steady-state density matrix $\rho_{ss}$ in the eigenbasis of the Hamiltonian $H_2$. It is evident that at $\alpha=0$ in the dissipative phase, the steady state mainly populates high-energy localized eigenstates [Fig. \ref{S2}(a)], whereas at $\alpha=\pi$, the steady state predominantly concentrates on mid-energy critical eigenstates [Fig. \ref{S2}(b)]. Furthermore, as mentioned in the main text,
for arbitrary $\alpha$, we can determine the proportion of localized and critical eigenstates in the steady states. This is defined as  $P_{l}=\sum_{i}\rho_{ii}$ ($P_c=\sum_k\rho_{kk}$), where $i (k)$ represents the index of the localized (critical) eigenstates $|\psi_i\rangle$ ($|\psi_k\rangle$), as illustrated in Fig. \ref{S2}(c). As $\alpha$ varies from $0$ to $\pi$, the system's steady state transitions from being primarily comprised of localized eigenstates to being predominantly composed of critical eigenstates. This transition highlights the potential of dissipation to manipulate the critical-localized transition. 

\subsubsection {B. Dissipation induced extended-critical transition}
Besides anomalous MEs that distinguish between localized and critical states, there are also those that distinguish between extended and critical states. We now discuss the impact of dissipation on such MEs using a quasi-periodic model with power-law hopping as an example. The Hamiltonian of this model is denoted as~\cite{DengXLS}
\begin{equation}\label{H3}
	H_3=-J \sum_{i,j\neq i} \frac{1}{|i-j|^{\kappa}}\vert  i\rangle\langle j\vert + \Delta \sum_j \cos(2 \pi\omega j + \phi)\vert j\rangle\langle j\vert.
\end{equation}
This model possesses exact anomalous MEs separating extended and critical states. We set $J=1$, $\kappa=0.5$, $\Delta=0.5$, $\omega=\frac{\sqrt{5}-1}{2}$, and $\phi=0$. Sorting the eigenstates according to their corresponding eigenvalues from low to high, the eigenstates corresponding to $E_n$ with $n<\omega L$ are extended, while those with $n>\omega L$ are critical~\cite{DengXLS}, where $L$ is the system size.

\begin{figure}[h]
	\centering
	\includegraphics[width=0.95\textwidth]{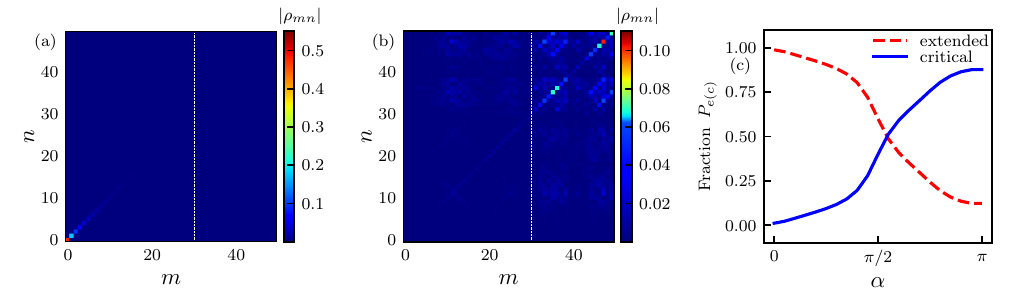}
	\caption{\label{S5} The magnitudes of the density matrix elements for steady states under dissipative phases (a) $\alpha = 0$ and (b) $\alpha = \pi$ in the eigenbasis of Hamiltonian given in Eq.~(\ref{H3}). Dashed lines are the MEs separating extended states from critical ones. (c) The fractions of extended eigenstates ($P_e$) and critical eigenstates ($P_c$) in steady states versus $\alpha$. Other parameters are $L=50$ and $\Gamma=1$.}
\end{figure}

Similarly, we introduce dissipative operators described by Eq. (4) in the main text with $l = 1$ and use open boundary conditions. Figs. \ref{S5} (a) and (b) illustrate the steady-state density matrix $\rho_{ss}$ distribution in the eigenbasis of the Hamiltonian $H_3$. It is apparent that in the dissipative phase at $\alpha = 0$, the steady state primarily occupies low-energy extended eigenstates, while at $\alpha = \pi$, the steady state predominantly concentrates on high-energy critical eigenstates. Fig. \ref{S5} (c) illustrates the fractions of extended eigenstates ($P_e$) and critical eigenstates ($P_c$) in steady states as a function of $\alpha$. We observe that as $\alpha$ varies from $0$ to $\pi$, the system's steady state evolves from being mainly constituted by extended eigenstates to being primarily composed of critical eigenstates. This suggests that dissipation can induce extended-critical transition.

\subsection{V. The impact of dissipation on the Dyson model}
\begin{figure}[h]
	\centering
	\includegraphics[width=0.9\textwidth]{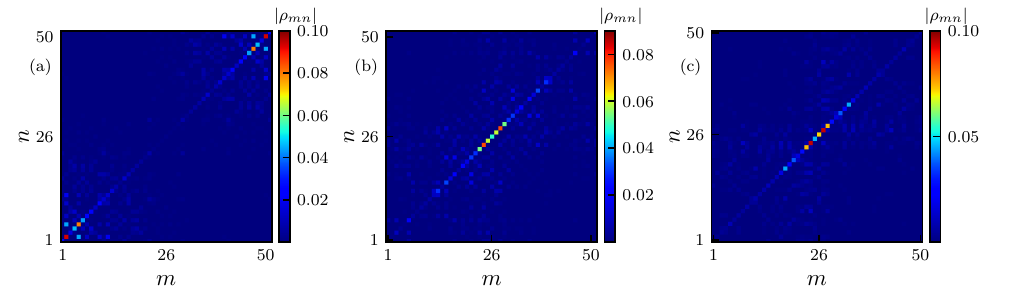}
	\caption{\label{S6} The absolute values of the steady-state density matrix elements with the dissipative phases (a) $\alpha = 0$, (b) $\alpha = \pi$ and (c) $\alpha=\pi$ in the eigenbasis of Hamiltonian given in Eq.~(\ref{DysonH}). The system size is $L=51$ for (a) and (b), and $L=50$ for (c). Here we fix $\Gamma=1$.}
\end{figure}

In the penultimate paragraph of the main text, we mentioned that our research could be used to prepare desired states and provided an example. Here, we provide another example by applying the dissipative operators we used to the Dyson model. This model includes random hopping, described as~\cite{DysonS}
\begin{equation}\label{DysonH}
	H_4=\sum_i J_i (\vert i\rangle\langle i+1\vert +\vert i+1\rangle\langle i\vert),
\end{equation}
where $J_i$ is the random number that follows the Poisson distribution $P(J_i=k)=\frac{\lambda^ke^{-\lambda}}{k!}$ with $\lambda$ being the mean of the random numbers, and we set $\lambda=3$ here. The localization length (proportional to $\ln|E|$) and the density of states of the Dyson model diverge as the eigenenergy approaches $0$.  We consider the impact of this dissipative operator Eq.~(4) in the main text with $l = 2$ on the Dyson model. We observe that when $\alpha=0$, the steady state mainly concentrates on the states at the upper and lower ends of the energy spectrum [Fig. \ref{S6}(a)], while when $\alpha=\pi$, the steady state predominantly consists of states in the middle of the energy spectrum [Figs. \ref{S6} (b) and (c)]. Therefore, regardless of the system's filling factor, we can prepare the system near $E=0$, thereby utilizing the intriguing properties of the states at $E=0$.

\begin{thebibliography}{36}
\bibitem{explainsign} We note that in the jump operator Eq. (\ref{jumpoperator}), there is no negative sign in front of both $\alpha$. It is easy to prove that when $\alpha=\pi/2$, the jump operator is Hermitian.
\bibitem{explain0} After the system reaches a steady state and dissipation is removed (i.e., $\Gamma$ becomes $0$), it is easy to obtain that the diagonal elements of $\rho_{mn}$ do not change with time. Therefore, the fraction of localized eigenstates ($P_l$) or extended eigenstates ($P_e$) remains constant over time. The off-diagonal elements evolve with time according to $\rho_{mn}(t) = e^{i(E_m-E_n)t}\rho_{mn}$, where $E_m$ and $E_n$ are the eigenvalues of the Hamiltonian $H$. When $\alpha$ is $0$ or $\pi$, $\rho_{mn}$ is very small when $E_m$ and $E_n$ are in the extended and localized regions, respectively. Furthermore, if we observe the system's dynamical behavior for a long period of time, the effects generated by off-diagonal elements will vanish. In our work, we focus on $P_l$ and $P_e$, thus concentrating only on the properties of the diagonal elements.
\bibitem{explain1} The disruption of localization discussed here diverges from the principles employed in previous studies. For example, many previous studies considered dephasing effects, and this disruption occurs by undermining the interference phenomenon crucial for AL. When dissipation is removed, the system's interference effects are restored, leading to the reoccurrence of AL. In our work, we disrupt localization by transforming particles from localized to extended states. After removing dissipation, particles stay in extended states, preventing the reappearance of localization.
\end{thebibliography}
\end{document}